\documentstyle[epsf,aps,twocolumn]{revtex}
\begin{document}
 
\title{ On the Improvement of Frequency Standards with Quantum Entanglement}

\author{ S. F. Huelga, C. Macchiavello\cite{ISI}, T. Pellizzari, and 
A. K. Ekert}
\address{Clarendon Laboratory, Department of Physics, University of
Oxford, Oxford OX1 3PU, U. K.}
\author{M. B. Plenio}
\address{Optics Section, The Blackett Laboratory, Imperial College,\\
London SW7 2BZ, U.K.}
\author{J. I. Cirac}
\address{Institut f{\"u}r Theoretische Physik, Universit{\"a}t Innnsbruck,\\
A-6020, Innsbruck, Austria}

\maketitle

\begin{abstract}
The optimal precision of frequency measurements in the 
presence of decoherence is discussed.
We analyse different preparations of $n$ two-level systems as well as 
different measurements procedures. We show that 
standard Ramsey 
spectroscopy on uncorrelated atoms and optimal measurements on maximally
entangled states provide the same resolution. The best resolution is 
achieved
using partially entangled preparations with a high degree of symmetry.
\end{abstract}
\section*{}

\narrowtext

The rapid development of laser cooling and trapping techniques has 
opened up new perspectives in high precision spectroscopy. 
Frequency standards based on laser cooled ions are expected
to achieve accuracies of the order of 1 part in $10^{14}-10^{18}$ 
\cite{wine:fs}.
In this letter we discuss the limits to the maximum precision achievable 
in the spectroscopy of $n$ two level atoms in the presence of decoherence.
This question is particularly timely in view of current efforts to 
improve high precision spectroscopy by means of quantum entanglement.

\begin{figure}
\epsfxsize8.0cm
\centerline{\epsfbox{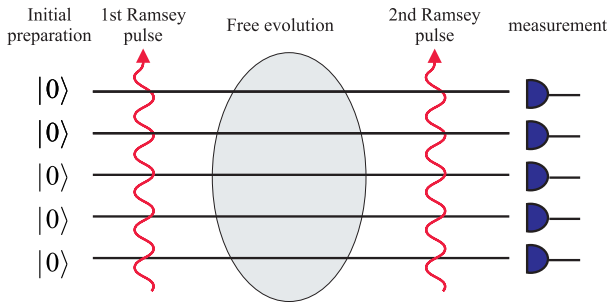} }
\vspace*{.2cm}
\caption{Schematic representation of Ramsey spectroscopy with 
uncorrelated particles.}
\end{figure}
In the present context standard Ramsey spectroscopy refers to the 
situation schematically depicted in Fig.~1.
An ion trap is loaded with $n$ ions initially prepared 
in the same internal state $|0\rangle$. A Ramsey pulse of frequency 
$\omega$ is 
applied to all ions. The pulse shape and duration are carefully 
chosen so that it drives 
the atomic transition
$|0\rangle\leftrightarrow |1\rangle$ of natural frequency
$\omega_0$ and
prepares an equally weighted superposition of the two internal states
$|0\rangle$ and $|1\rangle$ for each ion. Next the system evolves
freely for a time $t$ followed by
the second Ramsey pulse.
Finally, the internal state of each particle is measured. 
Provided that the duration of the Ramsey pulses is much smaller 
than the free evolution time $t$,
the probability that an ion is found in $|1\rangle$ is given by
\begin{equation}
        P = (1 + \cos \Delta\, t)/2.
\end{equation}
Here $\Delta=\omega-\omega_0$ denotes the detuning between the 
classical driving field and the atomic transition.

This basic scheme is repeated yielding a total duration $T$ of the 
experiment. 
The aim is to estimate $\Delta$ as accurately 
as possible for a given $T$ and a given number of ions $n$. 
The two quantities $T$ and $n$ are the physical resources 
we consider when 
comparing the performance of different schemes.
The statistical fluctuations
associated with a finite sample yield an uncertainty $\Delta P$ in the 
estimated value of $P$ given by
\begin{equation}
{\Delta P} = \sqrt{P (1-P)/ N}
\end{equation}
where $N=n T/t$ denotes the actual number of experimental data 
(we assume that $N$ is large). 
Hence the uncertainty in the estimated
value of $\omega_0$ is given by:
\begin{eqnarray} \label{shot-noise}
        {|  \delta \omega_0 | } = \frac{\sqrt{P (1-P)/N}}
        {|  dP/d\omega| }
                  && = {{1} \over {\sqrt{nTt}}}.
\end{eqnarray} 
This value is often referred to as the 
{\em shot noise limit} \cite{wine:unco}.

The theoretical possibility of overcoming this limit has been 
put forward recently
\cite{wine:corr1,wine:corr2}. 
The basic idea is to prepare 
the ions initially in an entangled state, which for
small $n$ seems to be practical in the near future.
To see the advantage of this approach, let us consider the case
of two ions prepared in the maximally entangled state \cite{remark2}
\begin{equation}
        |  \Psi \rangle = (|0 0 \rangle + | 11 \rangle)/\sqrt{2}.
\end{equation}
This state can be generated, for example, by the initial part of the network 
illustrated in 
Fig.~2. A Ramsey pulse on the first ion is followed by a controlled--NOT
gate \cite{c-not}.
After a free evolution period of time $t$ the state of 
the composite system in the interaction picture rotating at the driving 
frequency $\omega$ reads
\begin{equation}
        |  \Psi \rangle = (| 0 0 \rangle +  e^{-2 i \Delta t} \,| 11 \rangle)
	/\sqrt{2}\;.
\end{equation}
The second part of the network allows 
to disentangle the ions after the free evolution period. 
The population in state $|1\rangle$ of the first ion 
will now oscillate at a frequency $2 \Delta$
\begin{equation}
P_2 = (1 + \cos 2 \Delta t)/2.
\end{equation}
\begin{figure}
\epsfxsize8.0cm
\centerline{\epsfbox{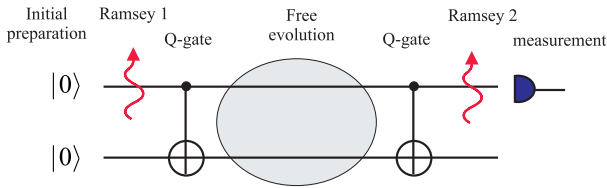} }
\vspace*{.2cm}
\label{fig2}
\caption{Spectroscopy with two maximally entangled particles.
The particles are entangled and disentangled by 
means of  ``controlled--NOT'' gates \protect\cite{c-not}.}
\end{figure}

This scheme can be easily generalised to the $n$ ion case 
by a sequence of controlled--NOT gates linking the first ion with
each of the remaining ones. In this way, a maximally entangled preparation 
of $n$ ions of the form
\begin{equation}
        |  \Psi \rangle = (|  00...0 \rangle + |  11...1 \rangle)/\sqrt{2}
\end{equation}
is generated. The final measurement on the first ion, after the free evolution
period and the second set of controlled--NOT gates, gives the signal
\begin{equation} \label{max-ent-signal}
        P_n = (1 + \cos n \Delta t)/2\;.
\end{equation}
The advantage of this scheme is that the oscillation frequency of the signal
is now amplified by a factor $n$ with respect to
the case of uncorrelated ions and the corresponding frequency uncertainty is 
\begin{equation}
{|  \delta \omega_0 | } = {{1} \over {n \sqrt{Tt}}}.
\end{equation}
Note that this result represents an improvement of a factor
 $1/\sqrt{n}$ over the shot noise limit (\ref{shot-noise}) by
using the same number of ions $n$ and the same total duration of the experiment
$T$ \cite{remark1} and it was argued 
that this is the best precision possible \cite{wine:rc}.

Let us now examine the same situation in a realistic experimental scenario,
where decoherence effects are inevitably  present. 
The main type of decoherence
in an ion trap is dephasing due to processes that cause random phase
changes while preserving the population in the atomic levels.
Important mechanisms that result in dephasing effects are
collisions, stray fields and laser instabilities.
We model the time evolution of the reduced density operator 
for a single ion $\rho$ in the presence of decoherence by the 
following master equation \cite{QuantumNoise}:
\begin{equation} \label{OptBloch}
\dot\rho(t)=i\Delta \left(\rho|1\rangle\langle1|
-|1\rangle\langle1|\rho\right) 
+ \gamma\left(\sigma_z\rho\sigma_z - \rho\right)
\label{master}
\end{equation}
%
the density matrix decay
[Actually, our analysis is not restricted to this particular model but 
holds for any process
where off-diagonal elements decay exponentially with time.]
Equation (\ref{master}) is written in a frame rotating at the frequency 
$\omega$. By $\sigma_z=|0\rangle\langle0|-|1\rangle\langle1|$ we denote 
a
Pauli spin operators.
Here we have introduced the decay rate $\gamma=1/\tau_{dec}$, where  
$\tau_{dec}$ is the decoherence time.
For the case of standard Ramsey 
spectroscopy this will give rise  to a broadening of signal (1):
\begin{equation}
        P = (1 + \cos \Delta\, t e^{-\gamma t})/2.
\end{equation}
As a consequence the corresponding uncertainty in the atomic frequency is 
no longer $\Delta$-independent. We now have

\begin{equation}
{|  \delta \omega_0 | } = \sqrt{{{1-\cos^2(\Delta t) e^{-2 \gamma t}} 
\over {n Tt e^{-2 \gamma t}}\sin^2(\Delta t)}}\;.
\end{equation}
In order to obtain the best precision it is necessary to optimise this 
expression as a function of the duration of each single measurement $t$. 
The minimal value is attained for
\begin{eqnarray}
\begin{array}{c}
\Delta t = k \pi/2 \,\,(k \;\; odd)\\
t = \tau_{dec}/2
\end{array}
\end{eqnarray}
provided that $T > \tau_{dec}/2$. Thus the minimum 
frequency uncertainty reads
\begin{equation}
{|  \delta \omega_0 | }_{opt} = \sqrt{{{2 \gamma e} \over {n T}}}
=\sqrt{{{2 e} \over {n \tau_{dec} T}}}.
\label{optunc}
\end{equation}

\begin{figure}
\epsfxsize8.0cm
\centerline{\epsfbox{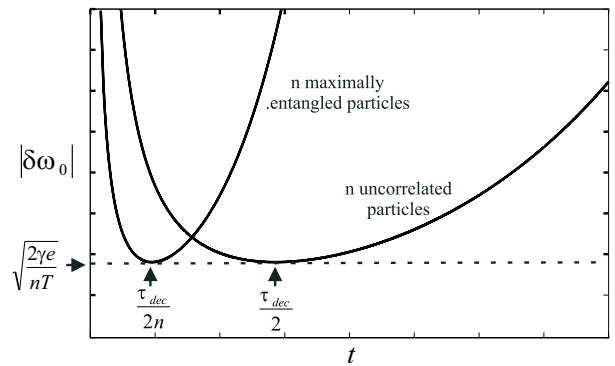} }
\vspace*{.2cm}
\caption{Frequency uncertainty $|\delta\omega_0|$ as a function of the 
duration
of a single shot $t$ for maximally entangled and uncorrelated particles.
Note that the minimum uncertainty is exactly the same for both 
configurations.}
\end{figure}

For maximally entangled  preparation the signal
(\ref{max-ent-signal}) in the presence of dephasing
is modified as follows:
\begin{equation}
        P_n = (1 + \cos n \Delta\, t e^{-n \gamma t})/2
\end{equation}
and the resulting uncertainty for the estimated value of the atomic 
frequency
is now minimal when
\begin{eqnarray}
\begin{array}{c}
\Delta t = k \pi/2n \,\,(k \;\; odd)\\
t = \tau_{dec}/2 n
\end{array}.
\end{eqnarray}
Interestingly, we recover exactly the same minimal uncertainty
as for standard Ramsey spectroscopy
(\ref{optunc}).
This effect is illustrated in Fig. 3. The modulus of
the frequency uncertainty $|\delta \omega_0|$ is plotted as
a function of the duration of each single experiment $t$ for 
standard Ramsey spectroscopy with $n$ uncorrelated particles and 
for a maximally entangled state with $n$ particles.
In the presence of decoherence
both preparations reach the same precision.  
This result can be intuitively understood by considering that maximally
entangled states are much more fragile in the presence of decoherence:
their decoherence time is reduced by a factor $n$ and therefore the duration 
of each single experiment $t$ has also to be reduced by the same amount.
Moreover, the limit (\ref{optunc}) represents the best  
accuracy for 
both uncorrelated and maximally entangled preparations and cannot be overcome 
by engineering different kinds of measurements as will be shown below. 

Note that the problem addressed in precision spectroscopy (i.e. the 
measurement of small atomic phase shifts) 
maps onto that of statistical distinguishability of nearby states, 
analyzed by Wootters \cite{bill:prd} and generalized by Braunstein 
and Caves \cite{caves1:prl,caves2:an}. 
They have provided an upper bound for the precision
in the estimation of a given variable that parametrizes 
a family of quantum states. In our case this variable is 
the detuning $\Delta$.
Moreover, the optimal measurements 
always correspond to a set of orthogonal projectors in
the $n$ ions Hilbert space.
It is worthwhile pointing out the generality of this result in the sense 
that it accounts 
for any possible joint measurement on the $n$ particles and  
any method of data analysis.
When the Braunstein and Caves optimization procedure is applied to 
either uncorrelated or maximally entangled preparations of $n$ ions
it yields the same limit (\ref{optunc}).

However, we will show in the following that with certain partially entangled
preparations one can overcome the limit  (\ref{optunc}).
Let us analyze first the case of generalized Ramsey spectroscopy, namely 
a scheme where the operator 
\begin{equation} 
S_{x} = \sum_{ k=1}^n \sigma_{x}^{k}
\label{ops}
\end{equation}
is measured after the free evolution period.
In (\ref{ops}) $\sigma_{x}=|1\rangle\langle0|+|0\rangle\langle1|$ denotes
the Pauli spin operator and the superscript $k$ refers to the $k$th 
particle.

We can easily evaluate the expectation values of the operators $S_x$ and
$S_x^2$ in terms 
of the corresponding quantities in the absence of decoherence:
\begin{equation}
\langle S_x \rangle \equiv \langle \sum_{k=1}^{n} \sigma_{x}^{k} 
\rangle = e^{-\gamma t} \langle S_x (\gamma=0) \rangle
\end{equation}
\begin{equation}
\langle S_{x}^2 \rangle = n + \langle \sum_{l \neq m} 
\sigma_{x}^l \sigma_{x}^m \rangle
= n + e^{-2 \gamma t}(\langle S_{x}^2(\gamma=0) \rangle - n).
\end{equation}

Finally, the resulting uncertainty in the atomic frequency is given by\\
\begin{equation}
{|  \delta \omega_o |} = \sqrt{{1 \over N} {{(\Delta S_x)^2} 
\over {\left({{\partial \langle S_x \rangle} 
\over {\partial \omega}}\right)^2}}}
\end{equation}
where $\langle\Delta S_x\rangle^2 = \langle S_{x}^2 \rangle - \langle S_x 
\rangle^2$ and $N$ denotes the total number of measurements
performed during the total time $T$. 
A straightforward calculation leads to an optimal duration of each 
measurement
$t_{opt}$ given by the solution of the following equation \cite{remark3}
\begin{equation}
n[1+(2\gamma t-1)e^{2\gamma t}]=\langle\Delta S_y(t=0)\rangle^2
\end{equation}
and the corresponding sensitivity takes the form
\begin{equation}
{|  \delta \omega_o | _{opt}} = \sqrt{{{2 n \gamma e^{2 \gamma t_{opt}}} 
\over {T \langle S_{x}(t=0) \rangle^2 }}}
\end{equation}
where $\langle S_{x}(t=0)\rangle$ and $\langle S_{y}(t=0) \rangle^2$ 
refer to the initial state preparation. 
The subscript {\em opt}
emphasizes that optimization with respect to $\Delta t$ has already
been taken into account, the minimum value being achieved for $\Delta
t = \pi/2$. Notice that each initial preparation is optimized for different 
values of the single shot time.
We can now state a lower bound for the precision
attainable within this approach as follows:
\begin{equation}
{|  \delta \omega_o | _{opt}} \geq \sqrt{{{2 n \gamma} \over {T
\langle S_{x}(0) \rangle^2}}} \geq \sqrt{{{2} \over {n \tau_{dec} T}}}.
\label{opt}
\end{equation}
Compared with the results above, a
maximum improvement of $1/\sqrt{e}$ in the resolution 
can be achieved.
Thus we have found that the bound ({\ref{optunc}) can be overcome for certain 
partially entangled states \cite{fam2}. 

We now analyze the best precision that can be achieved
when optimizing the experiment with respect to both the initial state 
preparation and the final measurement. 
The problem has been studied by means of a numerical optimization
procedure and we have restricted ourselves to
small numbers of ions in the trap.
The initial state preparation which leads to the best precision is of the form
\begin{equation}
|\psi_n\rangle=\sum_{k=0}^{\lfloor\frac{n}{2}\rfloor}a_k|k\rangle\;,
\end{equation}
where $|k\rangle$ denotes an equally weighted superposition of all states 
of $n$
ions which contain either a number $k$ or a number $n-k$ of excited states.
By $\lfloor\cdots\rfloor$ we denote the corresponding integer part.
The coefficients $a_k$ can be chosen to be real.
For example, the corresponding family of states for $n=4$ reads
\begin{eqnarray}
&&|\psi_4\rangle=a_0(|0000\rangle+|1111\rangle)+a_1(|0001\rangle+|0010\rangle\\
\;\;&&+|0100\rangle+|1000\rangle
+|1110\rangle+|1101\rangle+|1011\rangle+|0111\rangle)\nonumber\\
\;\;&&+a_2(|0011\rangle+|0101\rangle+|1001\rangle+|1100\rangle
+|1010\rangle+|0110\rangle)\nonumber\;.
\end{eqnarray} 
\begin{figure}
\epsfxsize8.0cm
\centerline{\epsfbox{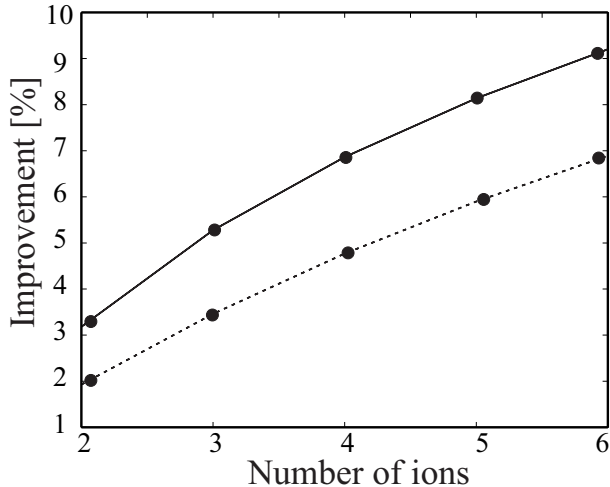} }
\vspace*{.2cm}
\caption{
The optimum percentual improvement in the precision
relative to the limit (\protect\ref{optunc}) as
a function of the number of ions $n$. Solid line: 
Numerical optimization with respect to the initial preparation and
application of the Braunstein and Caves algorithm for determinig the
optimal measurement. Dashed line: Optimized initial preparation 
and generalized Ramsey spectroscopy as the final measurement.
}
\end{figure}

Note that this family of states exhibits a high degree of symmetry:
it is completely symmetric under permutations of the $n$ ions and
under exchange of the excited and the ground state for each ion. 
The optimum percentual improvement in the precision 
relative to the limit (\ref{optunc}) as
a function of the number of ions $n$ is shown in Fig.~4. 
The solid curve
shows the improvement obtained by optimizing both the inital preparation
and the final measurement using the algorithm of Braunstein and Caves.
The dashed line exhibits the improvement obtained by optimizing only the 
initial preparation and performing the measurement given in Eq.~(\ref{ops})
corresponding to generalized Ramsey spectroscopy. The improvement obtained
by optimising the measurement is rather small. 
The question whether the (dashed) curve corresponding to Ramsey spectroscopy 
asymptotically 
saturates at the theoretical limit (\ref{opt}) or below remains to be
addressed. Moreover, whether the curve corresponding to Braunstein and Caves
optimization saturates at the same value or higher than the Ramsey curve
is an open question. 

%


In conclusion, we can state that
standard Ramsey spectroscopy is optimal for uncorrelated particles both 
in the
presence and in the absence of decoherence effects.
The use of maximally entangled states does not provide higher 
resolution as compared to using independent particles when decoherence 
is
present. The best sensitivity is achieved when the ions 
are initially prepared in highly symmetric but only partially entangled 
states.

We are grateful to A. Barenco, L. Cutler, G.M. D'Ariano, R. Derka, 
C.A. Fuchs, 
H.J. Kimble, P.L. Knight, H. Mabuchi, A. Steane, S. Williams, D.J. Wineland 
and 
P. Zoller for stimulating discussions.

This work was supported in part by the European TMR Research 
Network ERB 4061PL95-1412, 
the European TMR Research Network on Cavity QED,
the Royal Society, Hewlett-Packard, Elsag-Bailey and the 
Alexander von Humbold Stiftung.
SFH acknowledges support from  by DGICYT Project No. PB-95-0594 (Spain).
CM is supported by a TMR ``Marie Curie'' Fellowship.
TP holds an ``Erwin--Schr{\"o}dinger'' scholarship granted by the 
Austrian Science Fund. We also acknowledge private support from Otto Pellizzari.


\begin{thebibliography}{16}

\bibitem[*]{ISI} Also at: I.S.I. Foundation, Villa Gualino, V.le Settimio Severo 63,
10133 Torino, Italy.
 
\bibitem{wine:fs}
D. J. Wineland et al.,
IEEE {\em Trans. on Ultrasonics, Ferroelectrics and Frequency Control} 
A{\bf 37}, 515 (1990)
%


\bibitem{wine:unco}
W. H. Itano et al.,
{Phys. Rev.} A{\bf 47}, 3554 (1993).
%

\bibitem{wine:corr1}
W. J. Wineland et al., {Phys. Rev.} A{\bf 46}, R6797 (1992).
%


\bibitem{wine:corr2}
D. J. Wineland et al., {Phys. Rev.} A{\bf 50}, 67 (1994).
%

\bibitem{remark2}
We note this particular example does not represent exactly the original
proposal by Wineland
{\it et al.} \cite{wine:corr1,wine:corr2}. We have modified it to fit
the presentation in the paper, however, the two schemes are equivalent
as far as the resulting precision is concerned.\\

\bibitem{c-not}
A. Barenco {\it et al.}, Phys. Rev. Lett. {\bf 74}, 4083 (1995).


\bibitem{remark1}
In reference \protect\cite{wine:corr1} also
certain partially entangled preparations are analyzed which 
yield an improvement over the shot noise.

\bibitem{wine:rc}
J. J. Bollinger et al., {Phys. Rev.} A{\bf 54}, R4649 (1996). 
%

\bibitem{QuantumNoise}
C. W. Gardiner, {\it Quantum Noise} (Springer--Verlag, Berlin 1991).

\bibitem{bill:prd}
W. K. Wooters,
{Phys. Rev.} D{\bf 23},357 (1981)
%

\bibitem{caves1:prl}
S. L. Braunstein and C. Caves,
{Phys. Rev. Lett.} {\bf 72}, 3439 (1994)
%

\bibitem{caves2:an}
S. L. Braunstein, C. Caves and G. J. Milburn,
{Annals of Physics} {\bf 247}, 135 (1996)
%


\bibitem{remark3}
{We have restricted ourselves to
the same preparations as the ones analyzed in
\cite{wine:corr1} where only states yielding a resonance 
curve symmetric about $\omega_0$ are considered.}


\bibitem{fam2} For example, for $n=2$ these states are of the form
$a(|00\rangle+|11\rangle)+b(|01\rangle+|01\rangle)$
with real coefficients $a$ and $b$. For $a=\sqrt{0.3}$ and $b=\sqrt{0.2}$ we
have a 2$\%$ improvement in the sensitivity with respect to (\ref{optunc}).

\end{thebibliography}
\end{document}